\documentclass[a4paper,11pt]{article}
\usepackage{jheppub} 

\usepackage[T1]{fontenc} 
\usepackage[utf8]{inputenc}

\newcommand\MeV{\ensuremath{\text{~MeV}}}
\newcommand\MeVc{\ensuremath{\text{~MeV}/\text{c}}}

\makeatletter
\gdef\@fpheader{FERMILAB-PUB-19-262-PPD\hfill{}June 5, 2019}
\makeatother

\title{A practical way to regularize unfolding of sharply varying
  spectra with low data statistics}

\author{Andrei Gaponenko}
\affiliation{Fermi National Accelerator Laboratory,\\ Batavia, IL, USA}

\emailAdd{gandr@fnal.gov}

\abstract{Unfolding is a well-established tool in particle physics.
  However, a naive application of the standard regularization
  techniques to unfold the momentum spectrum of protons
  ejected in the process of negative muon nuclear capture
  led to a result exhibiting unphysical artifacts.
  A finite data sample limited the range in which
  unfolding can be performed, thus introducing a cutoff.  A sharply
  falling ``true'' distribution led to low data statistics near
  the cutoff, which exacerbated the regularization bias and
  produced an unphysical spike in the resulting spectrum.
  An improved approach has been developed to address
  these issues and is illustrated using a toy model.
  The approach uses full Poisson likelihood of data,
  and produces a continuous, physically plausible, unfolded
  distribution.  The new technique has a broad applicability
  since spectra with similar features, such as sharply falling
  spectra, are common.}

\begin{document}
\maketitle
\flushbottom

\section{\label{sec:into}Introduction}

The procedure of extracting a ``truth level'' physics distribution
that can be directly compared to a theoretical model from measured
quantities affected by finite detector resolution is
called unfolding \cite{Cowan1998}.
The mathematical problem of
unfolding is known to be ill-posed: truth level spectra that are
significantly different from each other can map into detector
distributions that have only infinitesimally small differences
\cite{Blobel:1984ku,Cowan1998,Cowan:2002in,Prosper:2011zz}.  The best
possible unbiased solution of an unfolding problem would have an
unacceptably large variance
\cite{Cowan1998}.
It has been shown that approximate solutions to unfolding problems can
be obtained by using a regularization procedure
\cite{Tikhonov1963a,Tikhonov1963b,Phillips1962}, which reduces the
variance of the result at the price of introducing a bias.
Implementations of unfolding algorithms for particle physics
applications, such as RUN/TRUEE \cite{Milke:2012ve} and TUnfold
\cite{Schmitt:2012kp} exist.  However they are based on the Gaussian
approximation of the log-likelihood function, and regularized
unfolding using the complete Poisson likelihood is still listed in the
``ideas'' section in this year's conference talk
\cite{Schmitt:phystat2019}.

The current work was performed in the context of measuring momentum
spectrum of charged particles emitted in the process of negative muon
capture on atomic nuclei at rest \cite{twist-mucapture}.  The median
number of data entries in non-empty bins of a reconstructed
2-dimensional distribution was about 10, necessitating the use of
Poisson likelihood in the analysis.  The spectrum varied by
more than an order of magnitude in the unfolding region.  A
straightforward application of standard regularization techniques,
introduced in section \ref{sec:notation} to a toy model,
defined in section \ref{sec:toy},
yielded unfolded spectra with
undesirable artifacts, as described in section \ref{sec:spline}
below.  Section \ref{sec:soe} presents modifications to the
unfolding procedure that allowed us extract the result without
unphysical features.   Section \ref{sec:lcurve} discusses the
choice of regularization strength, and \ref{sec:conclusion} summarizes
the findings.

\section{\label{sec:notation}Regularized unfolding}

The formulation of the unfolding problem involves an experimental
observable $x$, truth level variable $y$ with unknown distribution
$f(y)$, which we would like to determine, and detector response $R$.
Both experimental observables and truth level variables are in general
multidimensional.  For example, in the capture measurement
\cite{twist-mucapture} truth level information comprises particle
species and its true momentum, while experimental observables include
measured track momentum and its range in the detector.

We consider the case when the experimental spectrum is binned.
Detector response $R_{i}(y)$ is the expectation value of the number of
reconstructed events in bin $i$ given a true event occurring at $y$.
It describes all the detector effects: acceptance, efficiency, and
resolution---but is independent of the physics spectrum that is being
measured.
Detector response is usually determined from a Monte-Carlo simulation,
which forces a discretization in the $y$ space:
$\int{R_i(y)f(y)\,dy}\longrightarrow \sum_j R_{ij} f_j$ where $f_i$ is the
integral of $f(y)$ over bin $j$.
The bin size in the $y$ space has to be much smaller than the
experimental resolution in order for the simulation-derived $R_{ij}$
to be independent of the particular truth level spectrum shape used in
the simulation.
Small bin size in $y$ leads to a large number of unknowns $f_j$.  This
large number of unknowns is purely technical and is not related to the
number of effective degrees of freedom of the problem, which scales
with the size of the dataset \cite{Panaretos:2011bxp}.
However it can make non-linear
numerical minimization not feasible.  To reduce the number of degrees
of freedom to a physically appropriate value one can approximate the
unknown functions with splines \cite{Blobel:1984ku}, as is illustrated
later in this paper.

The expected number of data events in bin $i$, $\mu_i$, can be written
as
\begin{equation}
  \mu_i = N_{\text{true}} \sum_{j} R_{ij} f_{j}  + b_i
  \label{eq:predicted}
\end{equation}
where $N_{\text{true}}$ is the true number of events of interest in
the dataset, and $b_i$ is the background contribution.  A maximum likelihood
estimator for $f_{j}$ is formed by minimizing
\begin{equation}
\label{eq:Poisson}
-\log{\mathcal{L}}(d|\mu\{f\})  = -\sum_{i} ( d_i \log\mu_i - \mu_i )
\end{equation}
where $d_i$ is the observed number of data events in bin $i$.  However
the unfolding problem is ill-posed and must be regularized to obtain a
useful solution.  Regularized unfolding can be performed by minimizing
a combination of the log likelihood of data and a regularization
functional $S\{f\}$ \cite{Tikhonov1963a,Tikhonov1963b,Phillips1962}.
\begin{equation}
\label{eq:regularization}
{\mathcal{F}} = -\log{\mathcal{L}}(d|\mu\{f\}) - \alpha S\{f\}
\end{equation}
where $\alpha$ is the regularization parameter.

A widely used Tikhonov \cite{Cowan1998,Tikhonov1963a,Tikhonov1963b,Phillips1962}
regularization imposes a ``smoothness''
requirement on the spectrum by penalizing the second derivative of the
solution.  It therefore biases the result towards a linear function.
Another well established regularization, the maximum entropy (or
``MaxEnt'') approach \cite{Cowan1998}, is based on the entropy of a
probability distribution \cite{Shannon:1948zz}:
\begin{equation}
\label{eq:reg-maxent}
S_{\text{MaxEnt}} = -\sum_j q_{j} \ln(q_{j}), \qquad q_{j}\equiv f_{j}/\sum_k f_{k}
\end{equation}
It biases unfolding result towards a constant.

Unfolding with Tikhonov regularization can be implemented in a
computationally efficient way when $\chi^2$ minimization is used.
However this advantage is lost when Poisson likelihood is needed.  On
the other hand, MaxEnt guarantees that the unfolded spectrum is
positive, as is required for a particle emission spectrum, whereas
Tikhonov with a large regularization strength $\alpha$ pulls the
solution towards a straight line, which can cause some of $f_j$ to be
negative.  The present work uses the MaxEnt regularization term.

\section{\label{sec:toy}Toy model}

\begin{figure}[tp]
\begin{center}
\includegraphics[width=12cm]{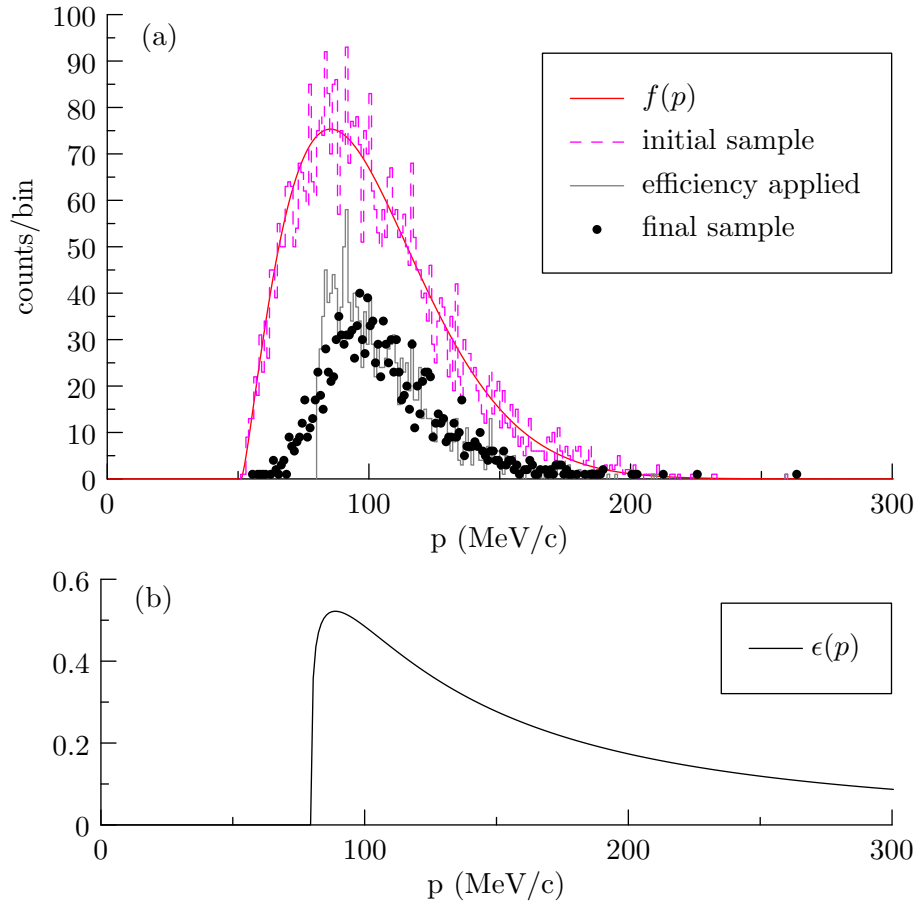}
\end{center}
\caption{\label{fig:toymodel}(a) Toy model momentum spectrum $f(p)$ and
  a random distribution of events drawn from it (initial sample),
  the distribution modified by detector acceptance times efficiency,
  and the final distribution after smearing.  See text for more details.
  (b) Toy model detector acceptance times efficiency vs momentum.}
\end{figure}

Unfolding issues will be illustrated using a one-dimensional toy model
that demonstrates some features first observed in the real life
application of the technique.
The model is based on the spectrum of protons ejected in the process
of negative muon nuclear capture.  The spectrum is known to follow an
exponential distribution in kinetic energy for large proton energies,
and to have a low energy threshold due to the Coulomb barrier
\cite{Measday:2001yr}.  We use the empirical functional shape and
parameters proposed in \cite{Hungerford:1999}, and convert the
distribution from kinetic energy to momentum space:
\begin{equation}
f(p) = C \frac{p}{\sqrt{p^2+m^2}}\times\left(1-\frac{1.40\MeV}{T(p)}\right)^{1.3279} \times\exp\left\{-\frac{T(p)}{3.1\MeV}\right\}
\label{eq:toy-truthp}
\end{equation}
where $m=938.27\MeV/c^2$ is the proton mass, $C$ is a normalization constant, and
$T(p)=\sqrt{p^2+m^2}-m$ is the kinetic energy of the proton.  The distribution
is shown in Fig.~\ref{fig:toymodel}(a).  Detector efficiency
times acceptance is modeled as
\begin{equation}
\epsilon(p)=\begin{cases}
0, & p <= p_0\\
\left(\displaystyle\frac{p}{p_0} - 1\right)^{0.2} \times\left(\displaystyle\frac{p}{p_0}\right)^2,
 & p > p_0
\end{cases}
\end{equation}
with $p_0=80\MeVc$, illustrated in Fig.~\ref{fig:toymodel}(b).  The
momentum resolution of the toy detector model as a Gaussian with
$\sigma=10\MeVc$.

A sample of 5000 momentum values was drawn from the $f(p)$
distribution (the ``initial sample'' in Fig.~\ref{fig:toymodel}(a)).
Some of the ``events'' were randomly dropped following the
$\epsilon(p)$ curve, then each remaining momentum smeared with the
Gaussian resolution to form the ``final sample'' of 1569 events used
for the unfolding tests below.
The response matrix for the tests was computed analytically and contains no
statistical fluctuations, corresponding to the limit of infinite MC
statistics.  The toy model contains no background.

\section{\label{sec:spline}Example of application}

\begin{figure}[tp]
\begin{center}
\includegraphics[width=\textwidth]{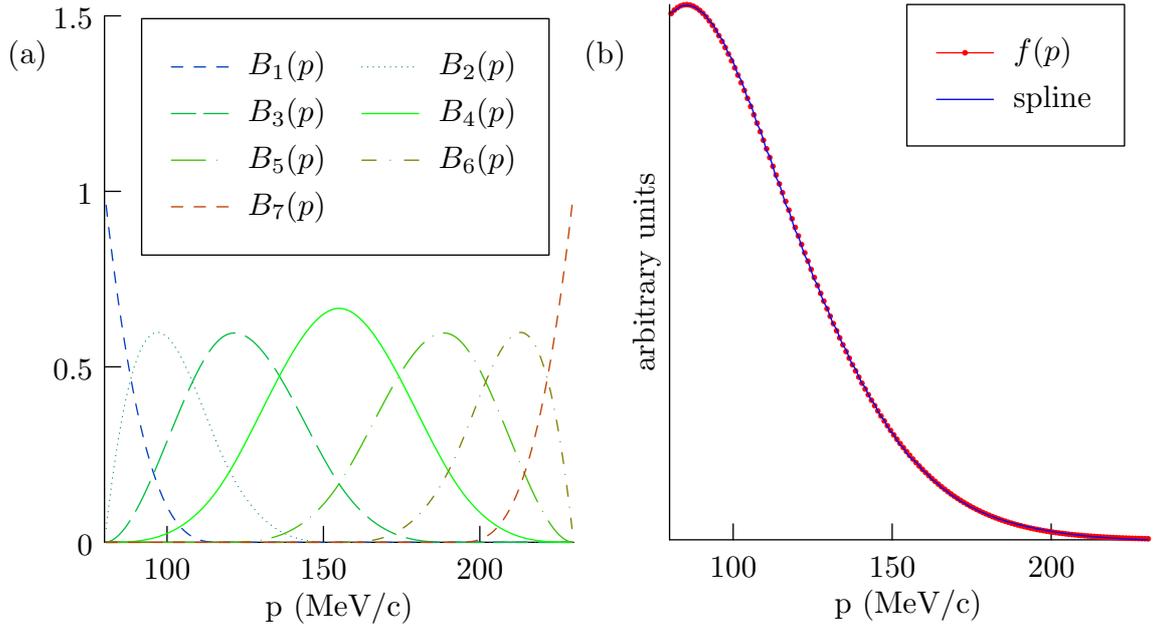}
\end{center}
\caption{\label{fig:genfit}(a) A set of B-splines.
  (b) $f(p)$ approximated by a linear combination of the splines.}
\end{figure}

To implement the approach outlined in section \ref{sec:notation} we
need to define an unfolding interval and select a set of splines on
that interval to approximate the distribution being unfolded.  Cubic
$B$-splines \cite{Boor1978b} provide a convenient basis for modeling
smooth continuous physics distributions.  Figure \ref{fig:genfit}(a)
shows a set of cubic $B$-splines obtained by placing 3 internal knots
that split the interval $80\MeVc<p<230\MeVc$ into 4 equal parts, and
locating all other necessary knots at the end points \cite{Boor1978b}.
Figure \ref{fig:genfit}(b) illustrates how a linear combination of
these splines can approximate the function $f(p)$ from
Eq.~\ref{eq:toy-truthp}:
\begin{equation}
\label{eq:splinesum}
f(p)\approx\sum w_i B_i(p).
\end{equation}
where $B_i(p)$ are the basis splines, and the $w_i$ are coefficients.

\begin{figure}[tp]
\includegraphics[width=\textwidth]{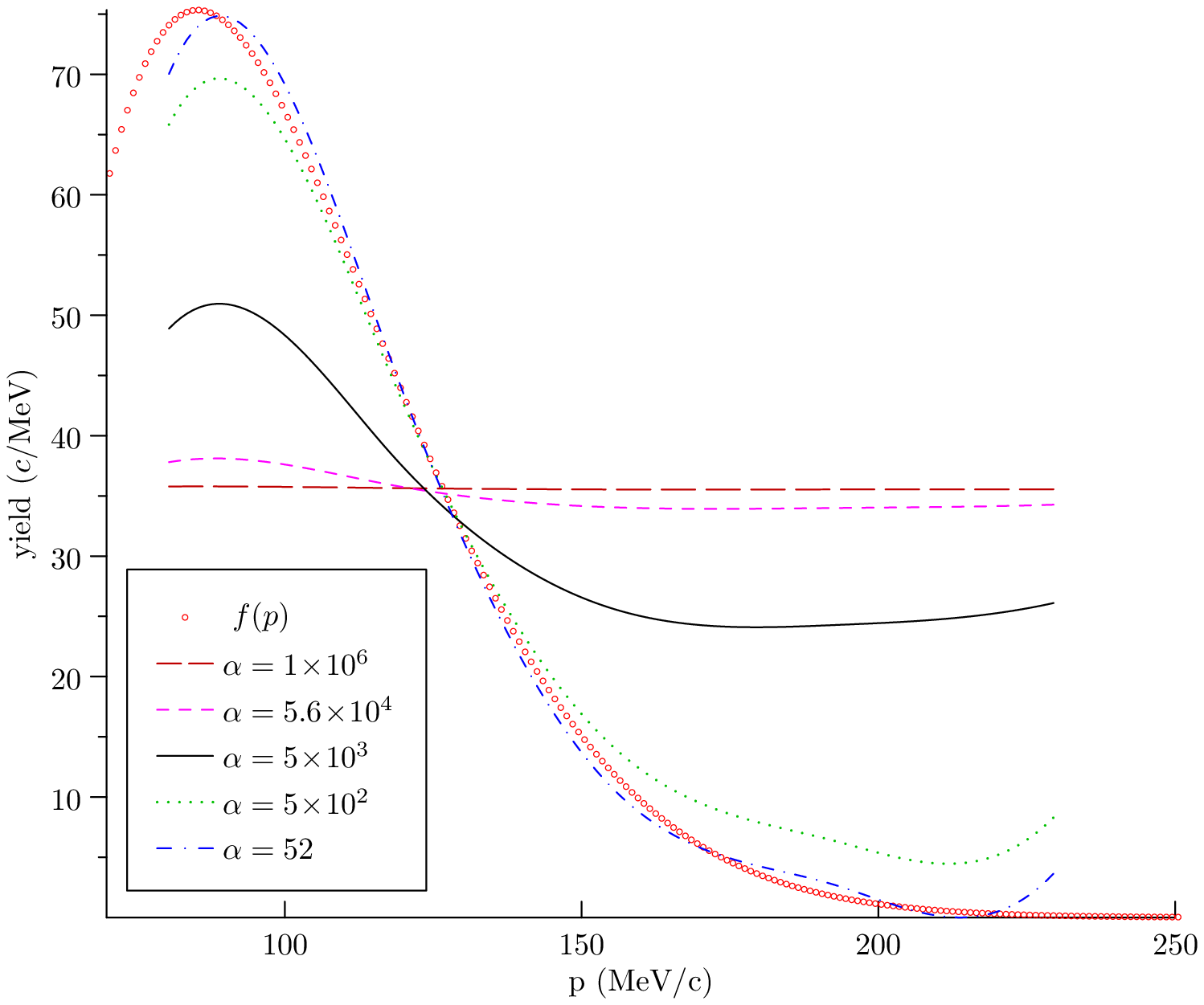}
\caption{\label{fig:result-spline}Unfolding results.}
\end{figure}

The unfolding is performed by minimizing Eq.~\ref{eq:regularization}
with respect to $w_i$  for a fixed value of $\alpha$.
The choice of a starting point is critical for the success of a
nonlinear multi-dimensional minimization.  Our implementation starts
with $f_j=\text{const}$ being an exact minimum of
(\ref{eq:regularization}) for $\alpha\to\infty$, and minimizes the
target functional for a large finite value of $\alpha$.  Then
$\log\alpha$ is reduced by a small amount, and the minimization is
re-run by using the previous minimum as the starting point.  As
$\log\alpha$ is further reduced, each new minimization starts at a
point that is linearly interpolated from the two previously found
mimima.  The process is repeated until the desired value of
regularization strength is reached.

Figure \ref{fig:result-spline} shows results for several settings of
regularization strength $\alpha$.  As it is reduced, the solution
changes from an almost constant function for $\alpha=10^6$, dominated
by the entropy term $S\{f\}$, to curves that are influenced by the
likelihood of ``data'' $\log{\mathcal{L}}(d|\mu\{f\})$.  The $f(p)$
spectrum used to produce the toy MC sample is also shown figure
\ref{fig:result-spline}.  One can see that $\alpha=5\times10^2$ is
still too large, and the corresponding curve does not reach $f(p)$ in
both its peak and tail regions.  On the other hand, it already
develops a unphysical rising behavior at the end of the unfolding
range.  Using a lower value $\alpha=52$ produces a spectrum that
oscillates about the ideal result and has a pronounced rise at the end
of the range.

The toy model example illustrates a typical behavior observed in a
real life applications of the unfolding technique.  In some cases the
procedure does not yield a satisfactory result for any value of
$\alpha$.  The result spikes at the end, and if one moves the upper
boundary of the unfolding interval the spike moves with it.
There are two effects that ``pull up'' the distribution at the end of
the unfolding region: the $S\{f\}$ regularization term, and the effect
of ``overflows'' (i.e.  reconstructed events that originated outside of
the unfolding interval).
The regularization term bias is exacerbated due to the fact that a
constant is not a good approximation for the rapidly falling true
distribution function.  A generalization of the MaxEnt approach, cross
entropy regularization \cite{Schmelling:1993cd,Cowan1998}, allows to
bias to an arbitrary reference distribution instead of a constant.
The distortion due to overflow events can be addressed by treating the
part of the signal distribution outside of the unfolding region as a
fixed shape background, as is done in e.g. \cite{Schmitt:2012kp}.  In
that approach the model of the signal distribution is not continuous,
because the resulting distribution in the unfolding region does not
generally match the a priori ``background'' distribution at the
interval ends.  Instead of trying to guess the steepness of the
``true'' distribution for the cross entropy and overflow background
priors, we suggest to fit it from data, as is detailed below.

\section{\label{sec:soe}An improved technique}

\begin{figure}[tp]
\includegraphics[width=\textwidth]{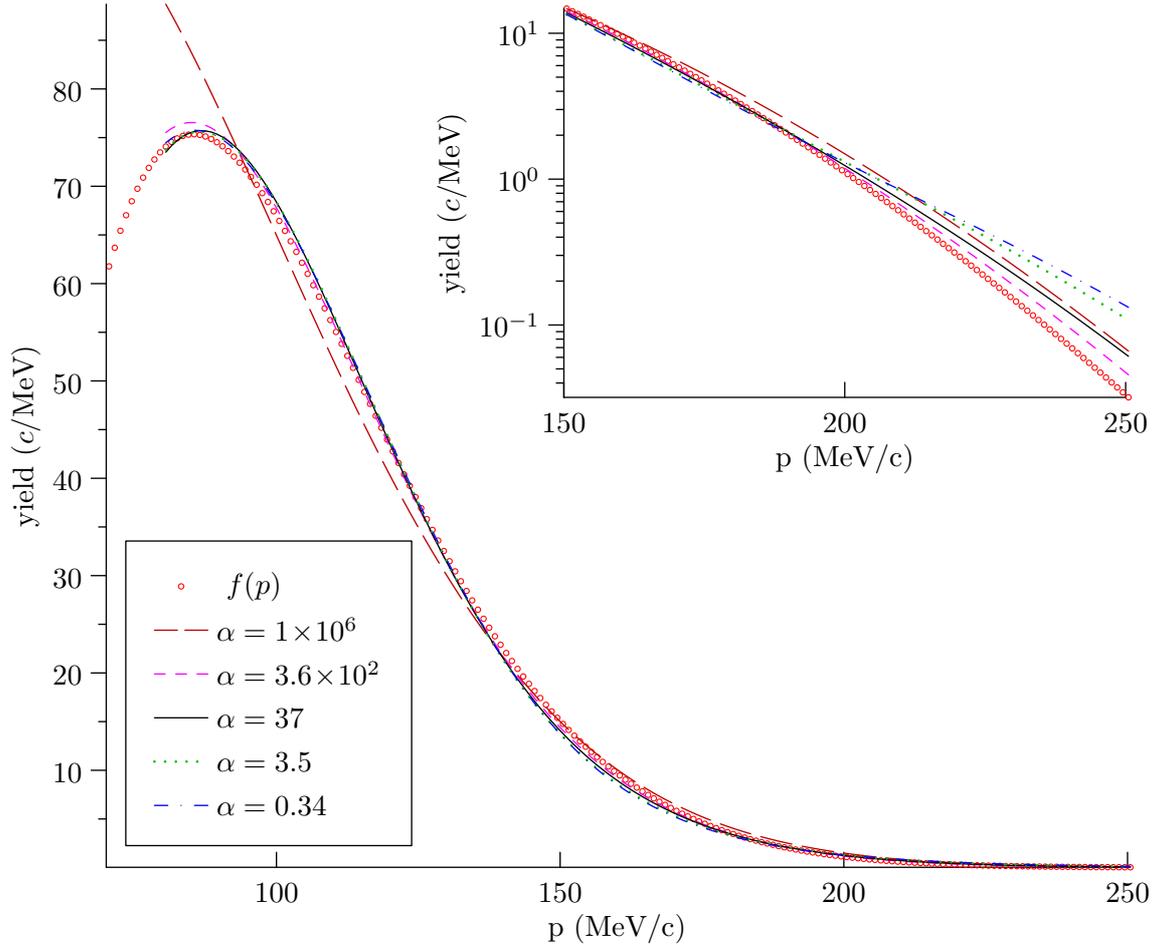}
\caption{\label{fig:result-soe}Results for the improved technique.}
\end{figure}

The main ideas to improve on the results of the previous section are:
\begin{itemize}

  \item Inside the unfolding region, bias towards a physically
    motivated function instead of a constant, with parameters of the
    function included in the fit.  For the spectrum of protons from
    muon capture example an exponential in kinetic energy was chosen,
    because the spectrum is know to approach this shape at high
    energies.  Note that the true distribution in the toy model
    (Eq.~\ref{eq:toy-truthp}) is not a simple exponential, however an
    exponential is a much better approximation for it in the unfolding
    interval than a constant.

  \item Include the ``overflow'' region in the minimization, and fit
    not just the normalization but also the exponential slope in that
    region.

  \item Require that the distribution is continuous and has two
    continuous derivatives.  This requirement connects the unfolded
    distribution to the overflow tail in a way that prevents the
    unphysical spike at the boundary.

\end{itemize}

Specifically, we represent
\begin{equation}
   f(p) = A \frac{p}{\sqrt{p^2+m^2}}\exp\{-\gamma T(p)\}
   \times
   \begin{cases}
     1 + \phi(p) & p_{\text{min}} < p \le p_{\text{max}} \\
     1 & p_{\text{max}} < p
   \end{cases}
\end{equation}
where $p_{\text{min}}$ and $p_{\text{max}}$ determine the limit
of the unfolding region, $m$ is the
mass of the particle and $T(p)$ its kinetic energy, $A$
and $\gamma$ are parameters pertaining to the exponential
behavior of the spectrum, and $\phi(p)$ is an arbitrary
function to be determined from the unfolding.
The regularization term has the form (\ref{eq:reg-maxent}) but now
acts on $1+\phi$ instead of $f$:
\begin{equation}
\label{eq:reg-maxent-final}
{S}_{\text{MaxEnt}} = -\sum_j \tilde{q}_{j} \ln(\tilde{q}_{j}),
\qquad \tilde{q}_{j}\equiv (1+ \phi_{j})/\sum_k (1+\phi_{k})
\end{equation}

The function $\phi(p)$ is approximated by a linear combinations of
cubic basis splines $B_l$ \cite{Boor1978b}
\begin{equation}
  \phi(p) = \sum_{l}^{n} w_{l} B_{l}(p),
  \qquad p_{\text{min}}<p\le p_{\text{max}}
\end{equation}
Here $w_{l}$ are the spline coefficients determined from the unfolding
process.  We require that the resulting spectrum has a continuous
second derivative, leading to
$\phi(p_{\text{max}})=\phi'(p_{\text{max}})=\phi''(p_{\text{max}})=0$,
which is provided by having a single-fold spline knot at the endpoint
$p_{\text{max}}$.  There are no continuity constraints at
$p_{\text{min}}$, therefore a 4-fold knot should be used at that point
to support the most general cubic spline shape.

To illustrate the modified technique, we use the same unfolding
interval $80\MeVc<p<230\MeVc$ as in section \ref{sec:spline} and the
same set of internal knots.  The resulting splines are $B_1$ to $B_4$
shown in Fig.~\ref{fig:genfit}.  Splines $B_5$ to $B_7$ would violate
the continuity condition and must not be included.  Like before, we
start with the maximally regularized solution and reduce $\log\alpha$
in small steps.  The resulting curves for several values of $\alpha$
are shown in Fig.~\ref{fig:result-soe}. Note that the starting
solution ($\alpha=1\times10^6$) is now close to an exponential, not a
constant, and that some of the resulting curves closely follow the
original $f(p)$ from Eq.~\ref{eq:toy-truthp}.

\section{\label{sec:lcurve}Choice of the regularization strength}
\begin{figure}[tp]
\includegraphics[width=\textwidth]{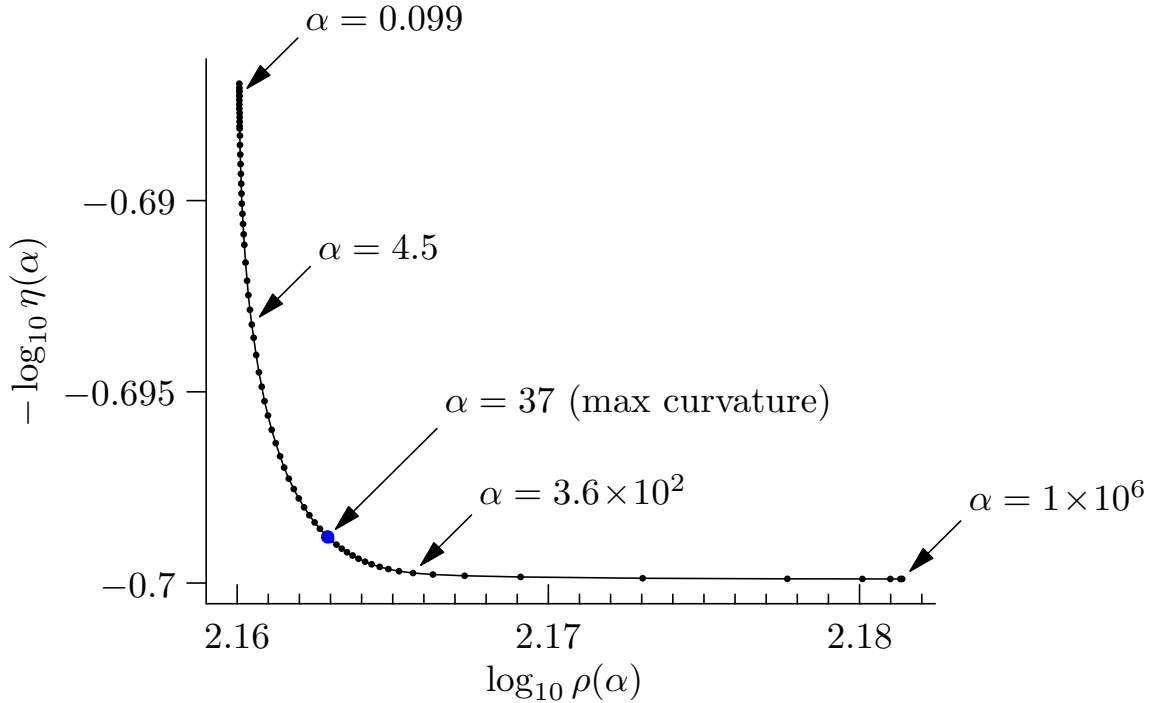}
\caption{\label{fig:lcurve-soe}L-curve for the improved technique.}
\end{figure}

The regularization strength $\alpha$ in Eq.~(\ref{eq:regularization})
should be chosen to provide an optimal balance between the variance
and the bias of the result.  The L-curve \cite{Hansen:1992,Hansen:1993}
provides a way to visualize a transition from strongly biased,
regularization term dominated solutions for large $\alpha$, to noise
dominated ones.  For a given $\alpha$ the minimization of
${\mathcal{F}}$ in Eq.~(\ref{eq:regularization}) yields particular
values of $\log{\mathcal{L}}$ and $S$.  Our code minimized a binned
likelihood ratio, so this is what we will use below instead of
the ``bare'' $\log\mathcal{L}$.  Following
\cite{Hansen:1993}, we define parametric functions
$\rho(\alpha)=-\log\left({\mathcal{L}}(d|\mu)/{\mathcal{L}}(d|d)\right)$
and $\eta(\alpha)=S$, and consider the curve $-\log\eta(\alpha)$ vs
$\log\rho(\alpha)$.  The choice of signs in the definition of the
$\eta$ term provides the conventional orientation of the ``L''.  A
plot of the curve is shown in Fig.~\ref{fig:lcurve-soe}.
As $\alpha$ is initially reduced from $\alpha=1\times10^6$, the curve
is almost horizontal, with quality of fit to data improving while not
significantly affecting the regularization term.  For small $\alpha$
the regularization penalty grows sharply without much improvement in
the data fit.  The optimal value of $\alpha$ lies in the transition
region, and can be defined as the point of the maximum curvature on
the L-curve \cite{Hansen:1993}.

In our example, the maximum curvature point is at $\alpha=37$.  The
corresponding unfolded spectrum is shown as the solid line in
Fig.~\ref{fig:result-soe}.  It is indeed a reasonable fit: the curves
for smaller $\alpha$ are farther away from the correct solution for
$p>200\MeVc$ and $150<p<170\MeVc$, while the curve for a larger
$\alpha=3.6\times10^2$ deviates more in the peak region
$p\approx80\MeVc$.

\section{\label{sec:conclusion}Conclusion}

The proposed method combines unfolding to an arbitrary function shape
in a phase space region with sufficient data statistics and a
parametric fit in the low statistics tail.  The whole distribution is
required to be twice continuously differentiable, which guarantees a
physically reasonable behavior of the result.  Factoring out the
exponential part of a sharply varying spectrum and applying the
regularization to just the deviation from the pure exponent reduces the
bias.  The use of the L-curve approach for finding the optimal
regularization strength has been demonstrated for Poisson likelihood
fit to data with the MaxEnt regularization term.

\acknowledgments

The author thanks Richard Mischke, Art Olin, Glen Marshall, Alexander
Grossheim, and Anthony Hillairet, who worked with me on the muon
capture analysis and provided encouragement vital for the completion
of this study.  In addition, Richard and Art provided valuable feedback
on the text of this article.

The numerical minimization code used for the study utilized the GNU
Scientific Library \cite{GSL}.  The figures were prepared with
Asymptote \cite{Asymptote}.

This document was prepared by the author using the resources of the
Fermi National Accelerator Laboratory (Fermilab), a U.S. Department of
Energy, Office of Science, HEP User Facility. Fermilab is managed by
Fermi Research Alliance, LLC (FRA), acting under Contract
No. DE-AC02-07CH11359.


\bibliographystyle{JHEP}
\bibliography{unfolding}

\end{document}